\documentclass[12pt]{iopart}
\usepackage{graphicx}
\usepackage{amssymb}
\begin{document}
\newcommand{\gguide}{{\it Preparing graphics for IOP journals}}

%

\newcommand{\eq}[1]{(\ref{#1})}
\newcommand{\pol}{\frac {d \sigma} {d \Omega}}
\newcommand{\bra}[1]{\langle #1 |}
\newcommand{\ket}[1]{| #1 \rangle}
\newcommand{\bm}[1]{\boldsymbol{#1}}
\newcommand{\B}{\mathcal{B}}
\newcommand{\T}{\mathcal{T}}
\newcommand{\R}{\mathcal{R}}
\newcommand{\thcm}{\theta_{\textrm{\footnotesize{c.m.}}}}
\newcommand{\tthcm}{\theta_{\textrm{\scriptsize{c.m.}}}}

\renewcommand{\d}{\mathrm{d}}
\renewcommand{\Re}{\textrm{Re}}
\renewcommand{\Im}{\textrm{Im}}

\title[]{Amplitude extraction in pseudoscalar-meson photoproduction: towards a situation of complete information}

\author{Jannes Nys, Tom Vrancx, Jan Ryckebusch}
\address{Department of Physics and Astronomy, Ghent University, Proeftuinstraat 86, B-9000 Gent, Belgium}
\ead{Jan.Ryckebusch@UGent.be}
\begin{abstract}
A complete set for pseudoscalar-meson photoproduction is a minimum set
of observables from which one can determine the underlying reaction
amplitudes unambiguously. The complete sets considered in this work
involve single- and double-polarization observables. It is argued that
for extracting amplitudes from data, the transversity representation
of the reaction amplitudes offers advantages over alternate
representations. It is shown that with the available
single-polarization data for the $ p \left( \gamma , K^+ \right)
\Lambda$ reaction, the energy and angular dependence of the moduli of
the normalized transversity amplitudes in the resonance region can be
determined to a fair accuracy. Determining the relative phases of the
amplitudes from double-polarization observables is far less evident.
\end{abstract}
%
\vspace{2pc}
\maketitle
\section{Introduction}
Figuring out how the interaction energy between the quarks comes about
at distance scales of the order of the nucleon size, is a topic of
intense research in hadron physics. Various classes of models, like
constituent-quark approaches, have been developed to understand the
structure and dynamics of hadrons in the low-energy regime of quantum
chromodynamics (QCD). While these models hold great promise, most of
them predict far more excited baryon states or ``resonances'' than
experimentally observed. The missing-resonance issue refers to the
fact that the currently extracted excitation-energy spectrum of
baryons contains less states than predicted in the available
models. This points towards a fundamental lack of understanding of the
underlying physics, which asks to be resolved.

Photoproduction of pseudoscalar mesons $M$ from the nucleon $N$, which
will be denoted as $N(\gamma,M)X$ throughout this work, continues to
be an invaluable source of information about the $N$ and $\Delta$
excitation spectrum and the operation of quantum chromodynamics (QCD)
in the non-perturbative regime \cite{aip1,aip2,aip3,aip4,aip5}. Whilst
most data are accumulated for the $N(\gamma,\pi)N$ reaction, recent
years have witnessed an enormous increase in the available data for
the $N(\gamma,\eta) N$, $N(\gamma,K)Y$ ($Y$ denotes hyperons),
$N(\gamma,2\pi)N$,\dots reaction channels. Partial-wave analyses of
these data in advanced coupled-channel models, have tremendously
improved our knowledge about resonances \cite{coup1, coup2, coup3},
but many properties of those are still not established. This asks for
a complementary approach to the extraction of the physical information
from the data for pseudoscalar-meson photoproduction. The experimental
determination of the reaction amplitudes represents a stringent test
of models, and may open up a new chapter in the understanding of
resonances' energy eigenvalues, decay and electromagnetic properties.

Thanks to technological advances and the concerted research efforts of
many groups, it is now feasible to measure several combinations of
$N(\gamma,M)X$ single- and double-polarization observables. Dedicated
research programs with polarized beams and targets are for example conducted at
the Mainz Microtron (MAMI) \cite{exp1}, the
Elektronen-Stretcher-Anlage (ELSA) in Bonn
\cite{exp2}, and the Continuous Electron
Beam Accelerator Facility (CEBAF) at Jefferson Lab \cite{exp3}.
It can be anticipated that these measurements involving polarization
degrees-of-freedom provide a novel window on the missing-resonance
problem and have the potential to constrain the theoretical models to
a large accuracy \cite{Dave2010, arenhov1998, lothar2014}.

Quantum mechanics dictates that measurable quantities can be expressed
in terms of combinations of bilinear products of matrix elements (=
complex numbers).
For $N(\gamma,M)X$, the kinematical conditions can be conveniently
determined by the combination of the invariant energy $W$ and by the
angle $\thcm$ under which the meson is
detected in the center-of-mass (c.m.)~frame. As we are dealing with a
time-reversal invariant process involving two spin-$\frac{1}{2}$
particles and one photon, the $N(\gamma,M)X$ observables are determined
in terms of seven real values (four complex numbers with one arbitrary
phase). A status of complete knowledge about $N(\gamma,M)X$ processes
over some kinematical range can be reached by mapping the kinematical
dependence of those seven real values. In the strict sense one defines
a ``complete set'' as a minimum set of measured $N(\gamma,M)X$
observables from which, at fixed kinematics, the seven real values can
be extracted unambiguously. In a seminal paper dating back to 1975,
Barker, Donnachie, and Storrow argued that a complete set requires
nine observables of a specific type \cite{hist1}. In 1996, this was
contested by Keaton and Workman \cite{hist2} and by Wen-Tai Chiang and
Tabakin \cite{chiang}. These authors asserted that eight well-chosen
observables suffice to unambiguously determine the four moduli and
three independent relative phases. In a recent paper \cite{TomComp}, we
have shown that for realistic experimental accuracies, the
availability of data for a complete set of observables does not
guarantee that one can determine the reaction amplitudes
unambiguously. The underlying reason for this is the fact that data
for polarization observables come with finite error bars and that
nonlinear equations connect the observables to the underlying
amplitudes. Accordingly, a status of complete quantum mechanical
knowledge about $N(\gamma,M)X$ appears to require more than eight
observables. In this paper we discuss some issues connected to the
extraction of the $N(\gamma,M)X$ reaction amplitudes from data.  This
procedure is known as ``amplitude analysis''. To date, the most
frequently used analysis technique of $N(\gamma,M)X$ data is
partial-wave analysis. Thereby, the
$\thcm$ dependence of the amplitudes
at fixed $W$ are expanded in terms of Legendre polynomials.

This paper is structured as follows. Section~\ref{sec:formalism}
summarizes the formalism that can be used to connect measured
$N(\gamma,M)X$ polarization observables to the reaction amplitudes in
the transversity basis. In Sec.~\ref{sec:results} a two-step procedure
is proposed for the purpose of extracting the reaction amplitudes from
data with finite error bars. Thereby, one first determines the moduli
of the amplitudes from single-polarization observables. Second, one
determines the phases of the reaction amplitudes from
double-polarization observables. We use real single-polarization data
for the $p \left( \gamma, K^+ \right) \Lambda$ reaction to illustrate
the procedure of extracting the amplitudes' moduli. Pseudo data with
realistic error bars are used to discuss the potential of determining
also the phases. Section~\ref{sec:conclusion} contains our conclusions
and prospects.

\begin{table}
\caption{\label{tab:transversity_representation} The connection
  between the $N(\gamma,M)X$ asymmetries of
  Eq.~(\ref{eq:asymmetry_definition}) and the normalized transversity
  amplitudes $a_k = r_k \exp \left( i \alpha _k \right)$. For the photon beam we
  adopt the following conventions: The $\B=+$ ($\B=-$) refers to a
  circularly polarized beam with positive (negative) photon
  helicity. The $\B= \pm \frac{\pi}{4}$ refers to oblique
  polarization, or photons which are linearly polarized along an axis
  tilted over an angle $\pm \frac{\pi}{4}$ with respect to the
  scattering plane.}

 \centering
\begin{tabular}{cccc}
\hline
Type & & $ \left( \B_1 , \T_1 , \R_1 \right) \quad \left( \B_2 , \T_2 , \R_2 \right)$ & \textbf{Transversity representation} \\
\hline
\noalign{\smallskip}\hline
Single & $\Sigma$  & $ (y,0,0)\quad (x,0,0)$ & $r_1^2 + r_2^2 - r_3^2 - r_4^2$ \\
& $ T $ & $ (0,+y,0) \quad (0,-y,0)$ & $r_1^2 - r_2^2 - r_3^2 + r_4^2$ \\
& $ P $ & $ (0,0,+y) \quad (0,0,-y)$ & $r_1^2 - r_2^2 + r_3^2 - r_4^2$ \\
\hline 
Double $\B \R$ & $C_x$ & $(+,0,+x) \quad (+,0,-x)$ & \phantom{$-$}$-2\Im(a_1a_4^* + a_2a_3^*)$\phantom{$-$}\\
& $C_z$ & $(+,0,+z) \quad (+,0,-z)$ & $+2\Re(a_1a_4^* - a_2a_3^*)$\\
& $O_x$ & $(+\frac{\pi}{4},0,+x) \quad (+\frac{\pi}{4},0,-x)$ & $+2\Re(a_1a_4^* + a_2a_3^*)$\\
& $O_z$ & $(+\frac{\pi}{4},0,+z) \quad (+\frac{\pi}{4},0,-z)$ & $+2\Im(a_1a_4^* - a_2a_3^*)$\\
\hline
Double $\B \T$ & $E$ & $(+,-z,0) \quad (+,+z,0)$ & $+2\Re(a_1a_3^* - a_2a_4^*)$\\
& $F$ & $(+,+x,0) \quad (+,-x,0)$ & $-2\Im(a_1a_3^* + a_2a_4^*)$\\
& $ G$ & $(+\frac{\pi}{4},+z,0) \quad (+\frac{\pi}{4},-z,0)$ & $-2\Im(a_1a_3^* - a_2a_4^*)$\\
& $H$ & $(+\frac{\pi}{4},+x,0) \quad (+\frac{\pi}{4},-x,0)$ & $+2\Re(a_1a_3^* + a_2a_4^*)$\\
\hline
Double $\T \R$ & $T_x$ & $(0,+x,+x) \quad (0,+x,-x)$ & $+2\Re(a_1a_2^* + a_3a_4^*)$\\
& $T_z$ & $(0,+x,+z) \quad (0,+x,-z)$ & $+2\Im(a_1a_2^* + a_3a_4^*)$\\
& $L_x$ & $(0,+z,+x) \quad (0,+z,-x)$ & $-2\Im(a_1a_2^* - a_3a_4^*)$\\
&$L_z$ & $(0,+z,+z) \quad (0,+z,-z)$ & $+2\Re(a_1a_2^* - a_3a_4^*)$\\
\hline
\end{tabular}
\end{table}

%
\section{Formalism}
\label{sec:formalism}
Any $N(\gamma,M)X$ observable can be written in terms of bilinear
products of four amplitudes 
$\left\{ \mathcal{M}_{1} , \mathcal{M}_{2}, \mathcal{M}_{3}, \mathcal{M}_{4}  \right\} $. The choices
made with regard to the spin-quantization axis for the $N,\gamma$ and $X$ 
determine the representation of the reaction amplitudes. Many
representations are available in literature and a detailed review is
given in Ref.~\cite{andy}. In Sec.~\ref{sec:results} we develop
arguments explaining that the use of transversity amplitudes offers
great advantages in the process of extracting the amplitudes from the
data.

We define the $z$-axis along the three-momentum $\vec{q}$ of the
impinging photon and the $xz$-plane as the reaction plane. The direction of the $x$-axis is such that the $x$-component of the meson momentum is positive. An alternate coordinate frame that is often used is the $x'y'z'$-frame, with $\vec{y}^{\,\prime} \equiv \vec{y}$ and the $z'$-axis pointing in the meson's three-momentum direction.
The \emph{transversity amplitudes} (TA) $b_i$ express the $\mathcal{M}_i$
in terms of the spinors $\ket{\pm}_y$ with a quantization axis
perpendicular to the reaction plane, and linear photon polarizations
along the $x$-axis or $y$-axis. The corresponding current operators are denoted by $J_x$ and $J_y$. One has 
\begin{equation}
\fl
b_1 =  {}_y\bra{+} J_y \ket{+}_y, \hspace{0.05\textwidth}
b_2 = {}_y\bra{-} J_y \ket{-}_y, \hspace{0.05\textwidth}
b_3 =  {}_y\bra{+} J_x \ket{-}_y, \hspace{0.05\textwidth}
b_4 =  {}_y\bra{-} J_x \ket{+}_y \; ,
\label{eq:transversityamplitudes}
\end{equation}
with $\bra{\pm}$ ($\ket{\pm}$) the recoil particle $X$ (target nucleon
$N$). The differential cross section for a given beam, target and recoil polarization $\B, \T$ and $ \R$ is denoted as
\begin{equation}
\frac{d \sigma}{d \Omega}^{(\B,\T,\R)} \; .\label{eq:cs_BTR}
\end{equation}
The notation ``$\B=0$'' refers to a measurement with an unpolarized beam. Similar definitions are adopted for the target and the recoil baryon. 
In computing the cross section \eq{eq:cs_BTR} for $\T = 0$, an
averaging over both target polarizations is implicitly assumed in
Eq.~(\ref{eq:cs_BTR}). Similarly, for $\R=0$ a summing over the two
possible recoil polarization states, is implicitly assumed. The unpolarized differential cross
section is given by
\begin{equation}
\frac{d \sigma}{d \Omega} = 
\frac{d \sigma}{d \Omega}^{(\B=0,\T=0,\R=0)} 
= \frac{K} {4} \sum_{i=1}^{i=4} \left| b_{i} \right| ^
    {2} \; ,
\label{eq:diffcs}
\end{equation}
with $K$ a kinematical factor.

The single- and double-polarization asymmetries $\mathcal{A}$ can be expressed as a ratio of cross sections:
\begin{equation}
\mathcal{A} = \frac{\pol ^{\left( \B_1,\T_1,\R_1 \right)} - 
\pol ^{\left( \B_2,\T_2,\R_2 \right)}}  
{\pol ^{\left(\B_1,\T_1,\R_1 \right)} + \pol^{\left( \B_2,\T_2,\R_2 \right)}} \; . 
\label{eq:asymmetry_definition}
\end{equation}
There are three single-polarization observables: the beam
asymmetry $\Sigma$ ($\B \neq 0, \T=0, \R=0$), the target asymmetry $T$
($\B=0, \T \neq 0, \R=0$), and the recoil asymmetry $P$ ($\B=0, \T=0, \R \neq 0$),
all contained in Table~\ref{tab:transversity_representation}. A
\emph{double} asymmetry, involves two polarized and one unpolarized
state. There are three types of double asymmetries: the target-recoil
asymmetries ``$\T \R$'' ($\B_1 = \B_2 = 0$), the beam-recoil ``$\B
\R$'' asymmetries ($\T_1 = \T_2 = 0$), and the beam-target ``$\B \T$''
asymmetries ($\R_1 = \R_2 = 0$). The definitions of the double asymmetries
are also contained in Table~\ref{tab:transversity_representation}. The double asymmetries involving a recoil polarization (denoted as `$A_{\{x,z\}}$') are often expressed in the $x'y'z'$-frame (denoted as `$A_{\{x',z'\}}$'). The $A_{\{x',z'\}}$ and $A_{\{x,z\}}$ are related through
\begin{equation}
\begin{array}{ll}
A_{x'} = A_x\cos\thcm - A_z\sin\thcm,\nonumber\\
A_{z'} = A_x\sin\thcm + A_z\cos\thcm,
\end{array}
\end{equation}

\begin{figure}
  \centering
\includegraphics[viewport=128 406 484 731, clip, width=0.75\textwidth]
{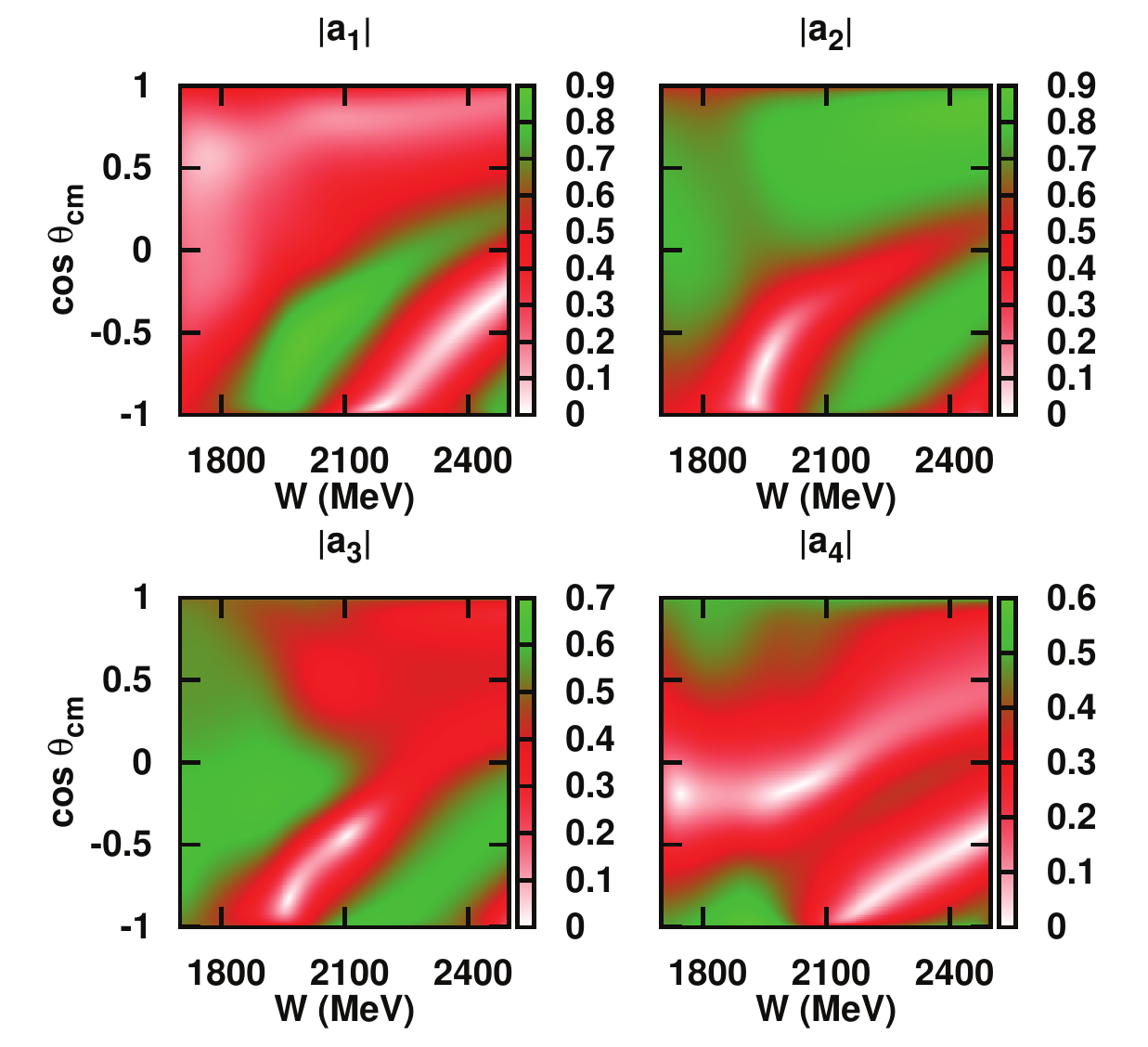}
\caption{The RPR-2011 predictions for the energy and kaon-angle dependence of the moduli $r_i= |a_i|$ of the NTA for the $p(\gamma,K^+)\Lambda$ reaction. }
\label{f1}
\end{figure}

The normalized transversity amplitudes (NTA) \cite{TomComp} are defined in
the following way:
\begin{equation}
a_{i}\equiv \frac{b_{i}}{\sqrt{\sum_{i=1}^{i=4} \left| b_{i} \right| ^
    {2}}} = 
\frac {b_{i}}
{\sqrt{\frac {4} {K}  
\frac {d \sigma} { d \Omega} } } \; .
\end{equation}
The NTA provide complete information about the $N(\gamma,M)X$
tranversity amplitudes $\left\{ b_1, b_2, b_3, b_4 \right\}$ after
measuring the differential cross section. Given the four TA, one can
determine the amplitudes in any quantization basis by means of a
unitary transformation \cite{TomComp,andy}. As can be appreciated from
Table~\ref{tab:transversity_representation}, any polarization
observable can be expressed in terms of linear and nonlinear equations
of bilinear products of the $a_{i}$. For given kinematics $ \left( W,
\thcm \right) $, the NTA $a_{k} \equiv
r_k \exp \left( i \alpha _k \right)$ are fully determined by six real
numbers conveniently expressed as three real moduli $r_{k=1,2,3} $ $
\; \biggl( r_4 = \sqrt{1 - \sum _{k=1} ^{k=3} r_k^2 } \biggr)$
and three real relative phases $\delta _{k=1,2,3} \equiv
\alpha_{k=1,2,3} - \alpha_4$. Only three phases are relevant as all
$N(\gamma,M)X$ observables are invariant under a transformation of the
type $\alpha _{k=1,2,3,4} \rightarrow \alpha _{k=1,2,3,4} + \beta$,
with $\beta \in \mathbb{R}$ an arbitrary overall phase. The moduli of
the NTA lie on a unit three-sphere in four dimensions:
\begin{equation}
r_1^2+r_2^2+r_3^2+r_4^2 = 1 \; .
\end{equation}  
Choosing $\alpha_4 =0 $, the NTA obey the equation of a unit six-sphere in seven dimensions \cite{Dave2010}: 
\begin{equation}
\left[ \Re \left( a_4 \right) \right] ^ 2 + \sum _{i=1} ^ {i=3} \left(
\left[ \Re \left( a_i \right) \right] ^ 2 + \left[ \Im \left( a_i
  \right) \right] ^ 2 \right) =1 \; .
\end{equation} 
Six real numbers are required to uniquely identify a point on a six-sphere. The NTA $a_i$, as defined in this work, and the Chew-Goldberger-Low-Nambu (CGLN) amplitudes $F_i$ are related through 
\begin{equation}
F_i = U_{ij}a_j\sqrt{\frac{d \sigma}{d \Omega}/\rho_0},
\end{equation}
with $\rho_0$ the ratio of the three-momenta of the meson and photon in the c.m.\ frame \cite{andy} and $U$ a unitary matrix given by \cite{TomComp}
\begin{equation}
U = -\frac{i}{\sin^2\thcm}
\left(
\begin{array}{*{4}{c}}
i \sin\thcm e^{i\tthcm} & i \sin\thcm e^{-i\tthcm} & 0 & 0 \\
i \sin\thcm & i \sin\thcm & 0 & 0 \\
-e^{i\tthcm} & e^{-i\tthcm} & e^{i\tthcm} & e^{-i\tthcm} \\
1 & -1 & -1 & -1
\end{array}
\right).
\end{equation}

In the remainder of this work, we consider the $ p (\gamma , K^ {+})
\Lambda$ reaction as a prototypical example of pseudoscalar-meson 
photoproduction from the nucleon. Despite the fact that the cross
section is several orders of magnitude smaller than those of pion
photoproduction, the selfanalyzing character of the $\Lambda$ hyperon is
an enormous asset for gathering a sufficiently large amount of
polarization observables and achieving a status of complete
information.

Before turning to an amplitude analysis of experimental data, in
Fig.~\ref{f1} we show model predictions for the energy and angular
dependence of the $ p ( \gamma, K^ {+} ) \Lambda$ normalized transversity amplitudes. We use the Regge-plus-Resonance (RPR) framework 
in its most recent version RPR-2011 \cite{LesleyPRC,LesleyPRL}. The
model has a Reggeized $t$-channel background and the $s$-channel
resonances $S_{11}(1535)$, $S_{11}(1650)$, $F_{15}(1680)$, $P_{13}(1720)$,
$P_{11}(1900)$, $D_{13}(1900)$, $P_{13}(1900)$,  and
$F_{15}(2000)$. The resonance content of the RPR-2011 model was
determined in a Bayesian analysis of the available data. The RPR
approach provides a low-parameter framework with predictive power for
$K^ {+}$ and $K^{0}$ photoproduction on the proton and the deuteron
\cite{PieterNPA}.

Interestingly, in Fig.\ref{f1} it is seen that the energy and angular
dependence of the moduli of the amplitudes $r_i = | a_{i} |$ display 
some interesting features. At forward kaon angles, where most the
strength resides, the $r_2$ plays a dominant role. The strongest
variations with energy are observed at backward kaon angles. No
dramatic changes in the energy dependence of the $| a_{i} |$ are
observed near the poles of the $s$-channel resonances. The smooth
energy dependence of the moduli at forward angles reflects the fact
that RPR-2011 predicts that the reaction under consideration receives
very important $t$-channel background contributions. The strongest
effects from the $s$-channel resonances are predicted to occur at
backward angles. In the forthcoming section it will be shown that
model predictions of the type shown in Fig.~\ref{f1} can already be
confronted with published data.

\begin{figure}
\centering
\includegraphics[viewport=51 453 545 763, clip, width=1.0\textwidth]
{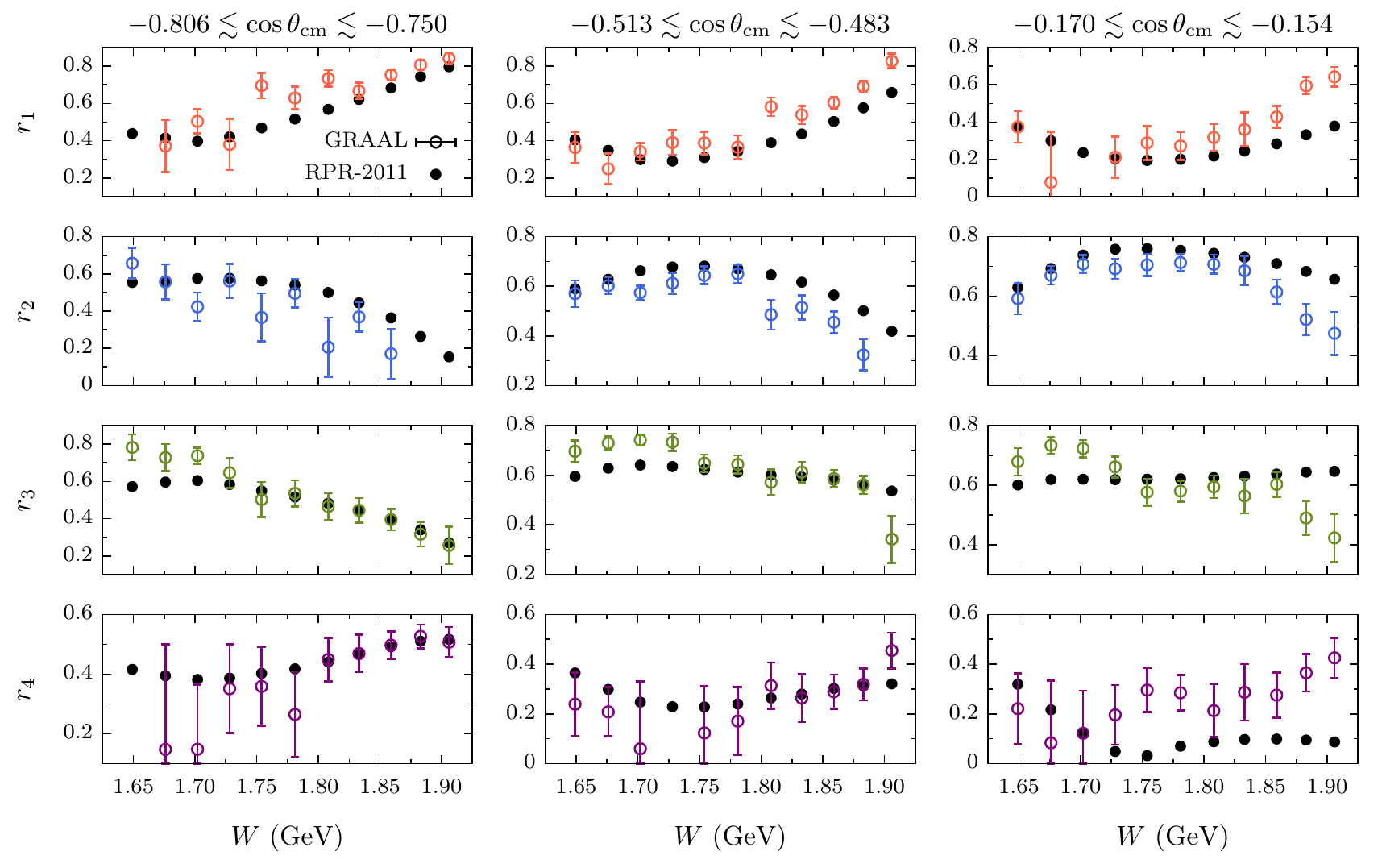}
\caption{The moduli $r_i$ for $p ( \gamma , K^+ ) \Lambda$ as a
  function of $W$ at three $\cos\thcm <0$ bins. The
  data are extracted from the single-polarization observables of
  Refs.~\protect \cite{GRAALresults1,GRAALresults2} with the
  Eq.~(\protect \ref{eq:1}). The dots are the RPR-2011 predictions. }
\label{f2}
\end{figure}

%
\section{Results}
\label{sec:results}
In Table~IV of Ref.~\cite{LesleyPRC} an overview of the published
experimental data for the $p ( \gamma , K ^{+} ) \Lambda$ reaction is
provided. Not surprisingly, in the quality and quantity of the
experimental results, there is a natural hierarchy whereby the
unpolarized data are most abundant. To date, the $p \left( \gamma , K
^{+} \right) \Lambda$ data base contains 4231 measured differential
cross sections, 2260 measured single-polarization observables (178
$\Sigma$, 2013 $P$, 69~$T$) and 452 data points for
double-polarization observables (320 $C_{(x,z)}$ and 132
$O_{(x',z')}$). For an amplitude analysis, it is important to realize
that the underlying physical realm is such that there are about 5
times as many $p \left( \gamma, K ^{+} \right) \Lambda$
single-polarization data than double-polarization data. In view of
this, the transversity representation of the amplitudes is a very
promising one. Indeed, in the transversity basis the
single-polarization observables are linked to the squared moduli of
the NTA by means of linear equations. From
Table~\ref{tab:transversity_representation} one easily finds that the squared NTA moduli can be directly obtained from 
\begin{equation}
\left\{
\begin{array}{ll}
r_1^2 = \frac{1}{4}\left(1 + \Sigma + T + P \right), \\ 
r_2^2 = \frac{1}{4}\left(1 + \Sigma - T - P \right), \\
r_3^2 = \frac{1}{4}\left(1 - \Sigma - T + P \right), \\ 
r_4^2 = \frac{1}{4}\left(1 - \Sigma + T - P \right) \; .
\end{array}
\right.
\label{eq:1}
\end{equation}
Accordingly, a measurement of $ \left( \Sigma, T, P \right)$ at given
$(W, \cos\thcm)$ provides good
prospects to infer the moduli $r_i (W,
\cos\thcm)$ of the NTA. We now
illustrate this with published data. The GRAAL collaboration
\cite{GRAALresults1, GRAALresults2} provides $p(\gamma, K^
     {+})\Lambda$ data for $ \left\{ \Sigma, T, P \right\}$ at 66 $(W,
     \cos\thcm)$ combinations in the
     ranges $1.65 \lesssim W \lesssim 1.91$~GeV (with a bin width of
     $\Delta W \approx 50$ MeV) and $-0.81 \lesssim
     \cos\thcm  \lesssim 0.86$ (with a
     bin width of $\Delta \cos\thcm
     \approx 0.3$).  Figure~\ref{f2} shows a selection of the
     extracted $r_i$ at three $\thcm$
     intervals in the backward kaon hemisphere along with the RPR-2011
     predictions. For 7 out of a total of 131 points, the $r_{i}$ could
     not be retrieved from the data. This occurs whenever one or more
     terms of the right-hand side of Eq.~(\ref{eq:1}) are negative due
     to finite experimental error bars. Note that there is negative
     correlation between the magnitude of $r_{i}$ and the size of the
     error bar $\Delta r_{i}$. Moreover, the smaller the $r_i$ the larger the
     probability that it cannot be retrieved from the $ \left\{
     \Sigma, T, P \right\}$ data.  From Fig.~\ref{f2} it is further
     clear that the RPR-2011 model offers a reasonable description of
     the magnitude and $W$ dependence of the extracted $r_{i}$. At
     forward angles (not shown here) the data confirm the predicted
     dominance of $r_{2}$ \cite{TomComp}. Another striking feature of
     Fig.~\ref{f2} is the rather soft energy and angular dependence of
     the moduli. As a matter of fact, the accomplished energy and
     angular resolution of the GRAAL data really suffices to map the
     moduli of the NTA in sufficient detail.
  
The above discussion illustrates that extraction of the moduli from
measured single-polarization observables can be successfully completed in
 over 90\% of the kinematic situations, given the present
accuracy of the data. Given information about values of the NTA moduli, inferring the NTA phases $\left\{ \delta _{1}, \delta_{2}, \delta _{3} \right\} $
from data requires measured double asymmetries. Some discrete ambiguities
arise from the nonlinear character of the equations that connect the
data to the underlying phases. We illustrate this with an example
using the complete set $\{{\Sigma, P, T}, C_x,O_x,E,F\}$. From the
expressions of Table~\ref{tab:transversity_representation} one readily finds
\cite{chiang,TomComp}
\begin{equation}
\label{eq:setforphases}
\left\{
\begin{array}{ll}
{r_1r_4}\sin\delta_1 + {r_2r_3}\sin\Delta_{23} = -\frac{C_x}{2}, \\
{r_1r_4}\cos\delta_1 + {r_2r_3}\cos\Delta_{23} = +\frac{O_x}{2},  \\
{r_1r_3}\cos\Delta_{13} - {r_2r_4}\cos\delta_2 = +\frac{E}{2}, \\
{r_1r_3}\sin\Delta_{13} + {r_2r_4}\sin\delta_2 = -\frac{F}{2} \; ,  
\end{array}
\right.
\end{equation}
where two independent $\delta_i$ and two dependent phases $\Delta _{ij}$ have been introduced:
\begin{equation}
\delta _i \equiv \alpha _i - \alpha _4 
\hspace{0.05\textwidth}
\textrm{and,} \hspace{0.05\textwidth}  \Delta _{ij} =
\delta _{i} - \delta_{j} \; .
\end{equation}
After determining the $r_i$, measured $\left\{ C_x, O_x \right\}$
allow one to compute $\left\{ \delta_1 , \Delta _{23} \right\}$, while
data for $\left\{ E, F \right\}$ yield $\left\{ \delta_2 , \Delta
_{13} \right\} $. Upon solving the above set for the cosine of a
specific angle one finds two solutions. The equation for the sine is
used to determine the quadrant in which each angle lies. This means
that upon solving the above set of four equations, there are four
possible combinations of angles $\left\{ \delta_1,
\delta_2,\Delta_{13}, \Delta_{23} \right \}$ which are compatible with
the double-polarization data given the moduli $ \left \{
r_1,r_2,r_3,r_4 \right \}$. We stress that in the transversity basis,
single-polarization observables are part of any complete set as they
provide the information about the NTA moduli. From the four possible
solutions for the $\left\{ \delta_1, \delta_2,\Delta_{13}, \Delta_{23}
\right \}$ of Eq.~(\ref{eq:setforphases}), the physical one can, in
principle, be selected from the trivial condition
\begin{equation}
\delta_1 -
\Delta_{13} - \delta_2 + \Delta_{23}  = 0 \; . 
\label{eq:condition}
\end{equation}
Several types of complete sets can be distinguished. Complete
sets necessarily involve double-polarization observables of two different kinds,
for example the combination of $\B \R$ and $\B \T$. In this work we
consider complete sets of the first kind and complete sets of the
second kind. The abovementioned set $\{{\Sigma, P, T}, C_x,O_x,E,F\}$
is a prototypical example of a complete set of the first kind with
four possible solutions to the phases. A similar procedure applied
to complete sets of the second kind, for example $\{{\Sigma, P, T},
C_x,O_x,E,H\}$, leads to eight solutions.

\begin{figure}
  \centering
  \includegraphics[viewport=73 406 527 745, clip, width=0.95\textwidth]
{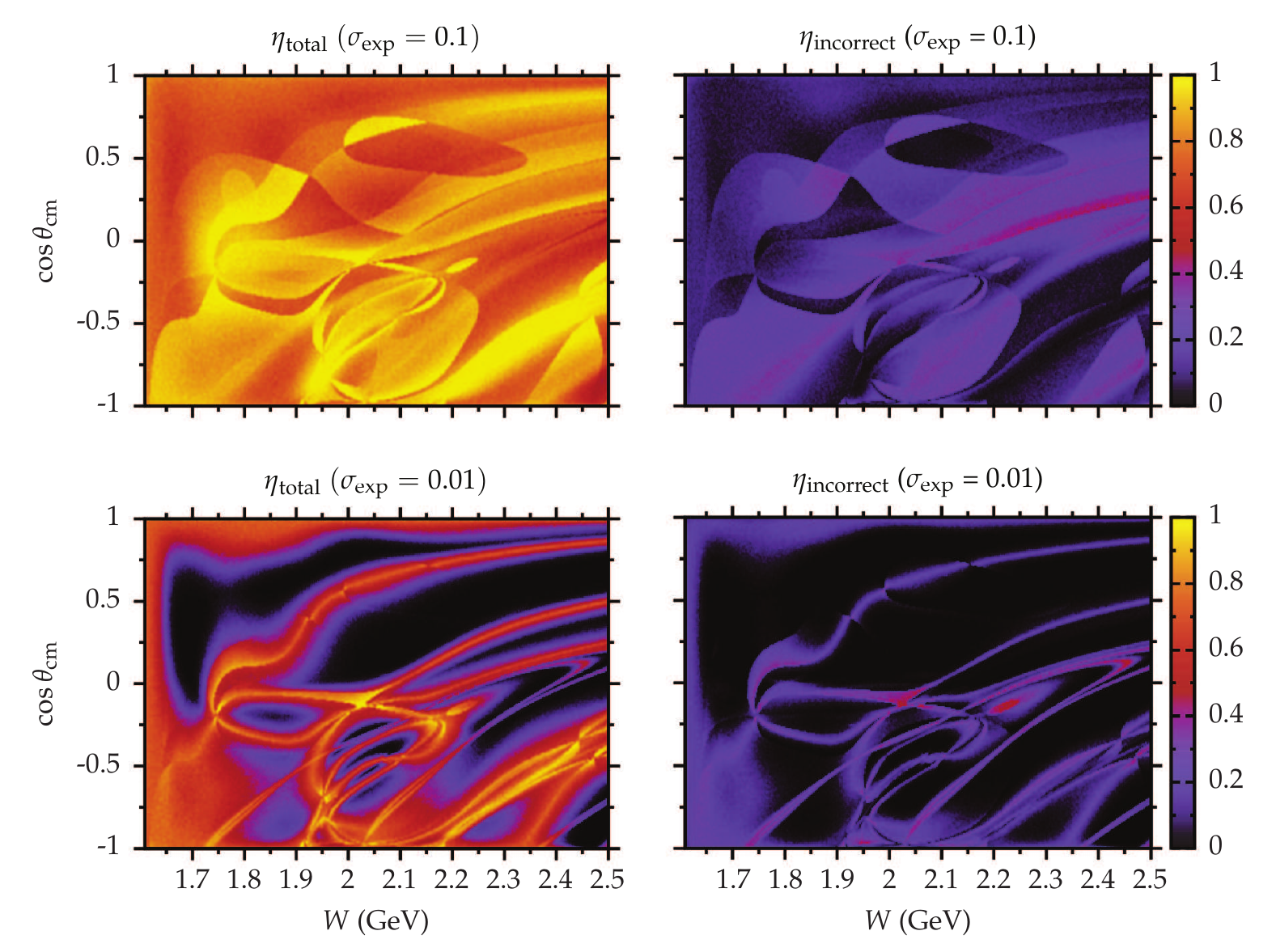}
\caption{The energy and angular dependence of the insolvabilities
  $\eta_{\textrm{\footnotesize{total}}}$
  ($\eta_{\textrm{\footnotesize{incorrect}}}$) at two values of the
  experimental accuracy. The pseudo data for the complete set
  $\{\Sigma, P, T, C_x,O_x,E,F\}$ are generated with the RPR-2011
  model. }
\label{fig:mapofeta}
\end{figure}

\begin{figure}
  \centering
  \includegraphics[viewport=84 451 509 735, clip, width=0.95\textwidth]
{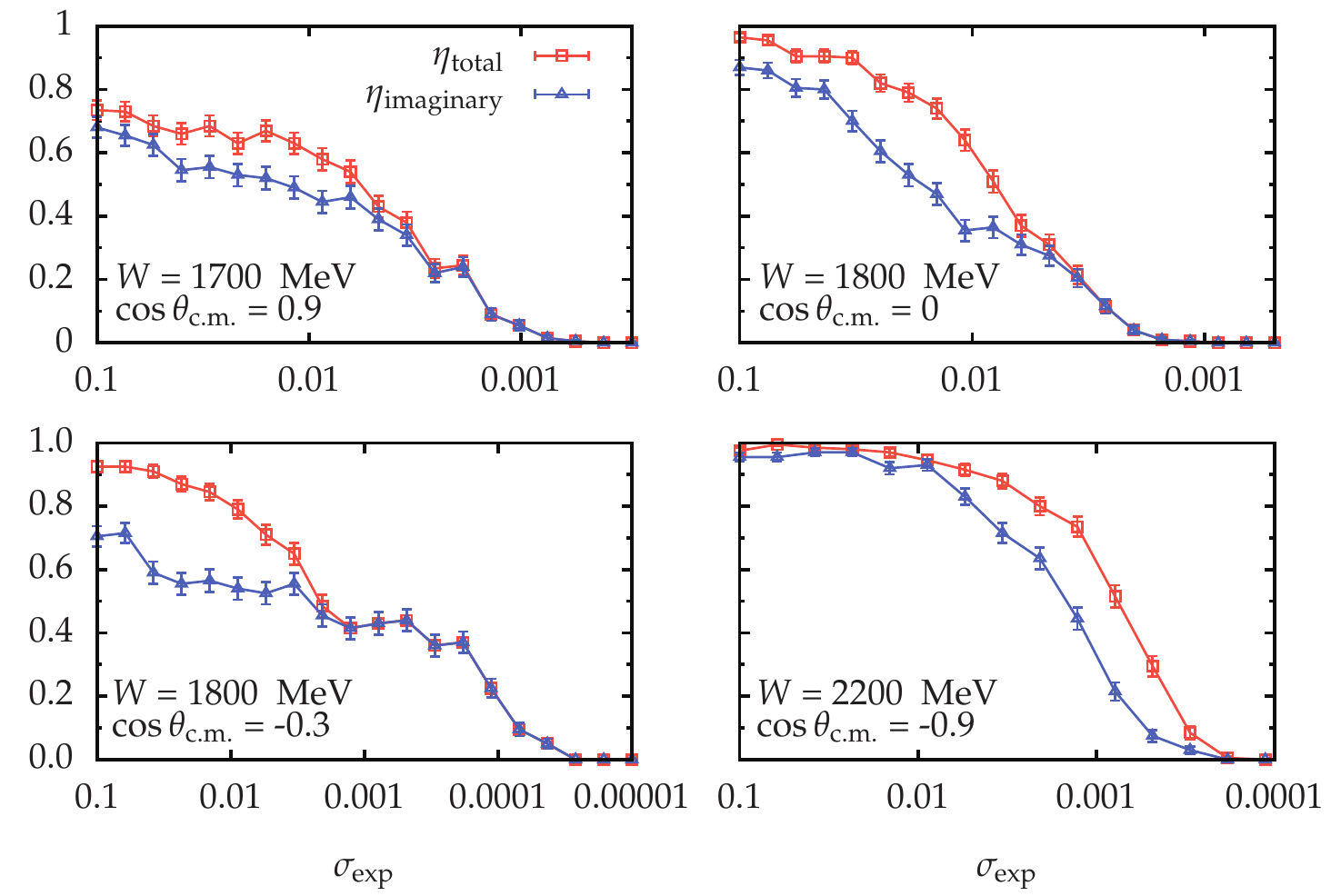}
\caption{The insolvability of the NTA as a function of the
  experimental accuracy at four different kinematical points. The data
  are generated from the complete set $\{\Sigma, P, T,
  C_x,O_x,E,F\}$. The squares (triangles) are the
  $\eta_{\textrm{\footnotesize{total}}}$
  ($\eta_{\textrm{\footnotesize{imaginary}}}$) insolvability.}
\label{fig:etaasfuncofexp}
\end{figure}

To date, the published double-polarization observables $\{C_{x}, C_{z},
O_{x'},O_{z'}\}$ for $p \left( \gamma, K^ {+} \right) \Lambda$ are of
the $\B \R$ type and do not comprise a complete set, as, to our
knowledge, neither $\B \T$ nor $\T \R$ data is available. In order to
assess the potential to retrieve the NTA phases from
double-polarization data with finite error bars, we have conducted
studies with pseudo data generated by the RPR-2011 model for $ p
\left( \gamma , K^ {+} \right) \Lambda$. We have considered ensembles
of 200 pseudo data sets for $\{\Sigma, P, T, C_x,O_x,E,F\}$ in a fine
grid of kinematic conditions. The pseudo data for all observables are
drawn from Gaussians with the RPR-2011 prediction as mean and a given
experimental accuracy $\sigma_{\textrm{\footnotesize{exp}}}$ determining the
standard deviation. The retrieved $\{ r_{i=1,2,3,4}, \delta
_{i=1,2,3} \}$ from analyzing the pseudo data with the aid of
Eqs.~(\ref{eq:1}), (\ref{eq:setforphases}) and (\ref{eq:condition}) do
not necessarily comply with the RPR-2011 input amplitudes, to which we
will refer as the ``actual solutions''. There are various sources of
error:
\begin{itemize}
\item[(i)] imaginary
solutions for the moduli upon solving the set of Eq.~(\ref{eq:1}) with input values of $\{ \Sigma, P, T \}$,
\item[(ii)] imaginary solutions for the phases upon solving a set of the type of Eq.~(\ref{eq:setforphases}) with input values for four different double-polarization observables,
\item[(iii)]
incorrect solutions which stem from the fact that the condition of Eq.~(\ref{eq:condition}) cannot be exactly obeyed
for data with finite errors. Thereby, the most likely solution does not necessarily coincide with the actual one.
\end{itemize}

\begin{figure}
  \centering
  \includegraphics[viewport=52 100 497 736, clip, width=0.85\textwidth]
{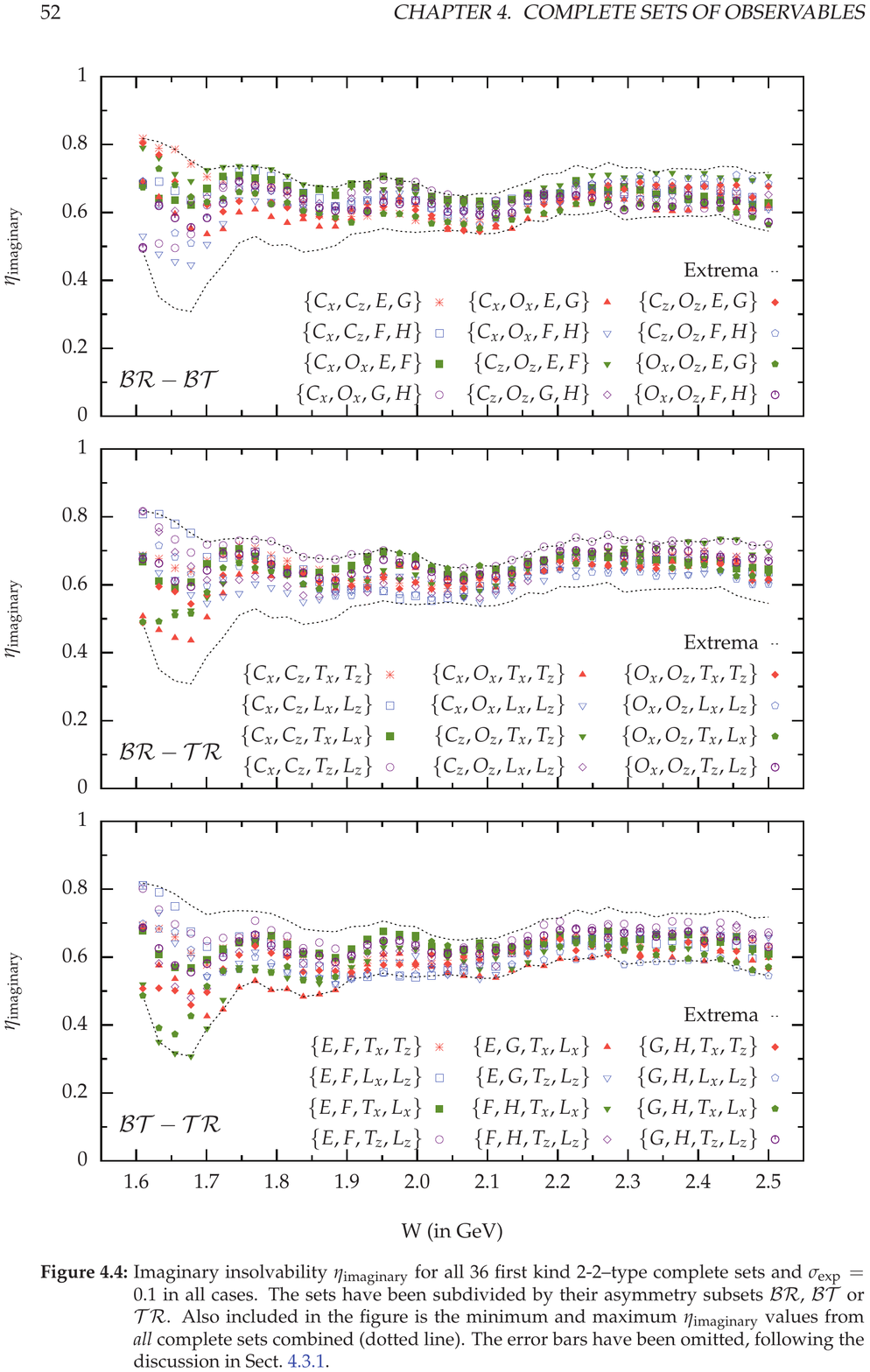}
\caption{The energy dependence of the $\eta_{\textrm{\footnotesize{imaginary}}}$ for $\sigma _{\textrm{\footnotesize{exp}}} =
  0.1$. We consider all 36 complete sets of the first kind of the $\B\R$,
  $\B\T$, and $\T \R$ type. Also shown is the minimum and maximum
  value of $\eta_{\textrm{\footnotesize{imaginary}}}$.}
\label{fcomparison}
\end{figure}

The insolvability $\eta$ at a specific 
kinematic point $\left( W , \cos \theta _{\textrm{\footnotesize{c.m.}}} \right)$ is introduced as the fraction of simulated complete
data sets that are solved incorrectly or have imaginary solutions:
\begin{equation}
 \eta_{\textrm{\footnotesize{total}}} = \eta_{\textrm{\footnotesize{incorrect}}} + \eta_{\textrm{\footnotesize{imaginary}}} \; . 
\end{equation}
In the simulations, we first check whether all values of the retrieved
NTA phases and moduli are real. Subsequently, we check whether those
extracted real values for $\{ r_{i=1,2,3,4}, \delta _{i=1,2,3} \}$
coincide with the ``actual' ones. As an example,
$\eta_{\textrm{\footnotesize{incorrect}}}=0.4$ indicates that $40\%$
of the evaluated sets of ``complete'' pseudo data that provide real
solutions, are incorrect. For,
$\eta_{\textrm{\footnotesize{total}}}=0$ the actual NTA can be
retrieved in all considered sets of pseudo data. Figure
\ref{fig:mapofeta} shows the angular and energy dependence of the
insolvabilities for the complete set $\{\Sigma, P, T, C_x,O_x, E, F
\}$. It can be concluded that for realistic experimental accuracies of
the order of 10\% of an asymmetry's maximum possible value
(corresponding with $\sigma _{\textrm{\footnotesize{exp}}}=0.1 $) the
$\eta_{\textrm{\footnotesize{total}}}$ are quite substantial and of
the order 0.6--0.7. Thereby, the largest contribution is from the
imaginary solutions upon determining the NTA phases from measured
double asymmetries. Figure \ref{fig:etaasfuncofexp} illustrates that
increasing the experimental resolution clearly improves the overall
solvability. In the limit of vanishing error bars, the complete sets
of observables do, indeed, allow one to retrieve the full information
about the NTA from the data. For achievable experimental accuracies
this is unfortunately not the case in the majority of situations.

 Although incorrect solutions make up the smaller contribution to
 $\eta_{\textrm{\footnotesize{total}}}$, they can never really be identified
 in an analysis of real data without invoking a model. Incorrect
 solutions originate from assigning the most likely solution as the
 actual solution, which is not a statistically sound procedure. A more
 conservative approach would consist of imposing a tolerance level on
 the confidence interval of the most likely solution. Then, the most
 likely solution would be accepted as the actual one if 
 it has a certain minimum statistical significance. As discussed at
 the end of Sec.\ IV C 2 in Ref.~\cite{TomComp}, however, imposing a
 tolerance confidence level would not be effective as the entire
 elimination of the incorrect solutions would lead to a rejection of
 the lion’s share of the fraction of correct solutions. This would
 result in an almost 100\% insolvability.

The above conclusions about the possibility to determine the NTA from
``complete'' sets are based on a statistical analysis of pseudo data
for $\{\Sigma, P, T, C_x,O_x, E, F \}$. We now address the question
whether there are large differences between the various complete sets
for retrieving the NTA. To this end, we have performed simulations for
all 36 complete sets of the first kind. We find that
$\eta_{\textrm{\footnotesize{imaginary}}}$ is consistently the most important source
of NTA extraction failure. The results of our simulations are
displayed in Fig.~\ref{fcomparison}. It can be concluded that for
$\sigma _{\textrm{\footnotesize{exp}}}=0.1$ and $W \gtrsim 1.8$~GeV, the $\eta_{\textrm{\footnotesize{imaginary}}}$ cluster
around 0.6-0.7 for all 36 complete sets of the first kind. Larger variations are observed in the threshold region. This means that for the bulk of the kinematic conditions, 
no combination of observables performs any better than
others. The results of Fig.~\ref{fcomparison} involve an averaging
over $\cos \theta _{\textrm{\footnotesize{c.m.}}}$.

\section{Conclusion}
\label{sec:conclusion}
We have addressed the issue of extracting the reaction amplitudes from
pseudoscalar-meson photoproduction experiments involving
single-polarization and double-polarization observables. A possible
roadmap for reaching a status of complete information in those
reactions has been sketched. We suggest that the use of transversity
amplitudes is tailored to the situation that experimental information
about the single-polarization observables $ \{ \Sigma, P, T \}$ is
more abundant (and most often more precise) than for the
double-polarization observables. Linear equations directly connect $\{
\Sigma, P, T \}$ to the squared moduli $r_{i=1,2,3,4}^2$ of the
normalized transversity amplitudes. From an analysis of available $p
(\gamma, K^+) \Lambda$ single-polarization data, we could extract the
$r_i$ in the majority of considered kinematic situations. Extracting
the NTA independent phases is far more challenging as the equations
which link them to the double asymmetries are nonlinear. Using studies
with pseudo data, we have found that for currently achievable experimental
accuracies, the actual phases cannot be extracted in
the majority of situations. Overcomplete sets which involve more than
four double asymmetries considerably improve on this figure of merit
and these are subject of current investigations.

\end{document}